\documentstyle[epsfig,elsart12]{elsart}

\begin{document}

\pagestyle{empty}
\begin{frontmatter}
\begin{flushright}
    FERMILAB Pub-97/226-E\\
\end{flushright}

\title{Scintillation Counters for the D\O\ Muon Upgrade}
\author[tat]{B.S. Acharya},
\author[fna]{B. Baldin},
\author[tat]{S. Banerjee},
\author[pan]{S.B. Beri},
\author[pan]{V. Bhatnagar},
\author[del]{M. Bhattacharjee},
\author[fna]{A. Brandt},
\author[msu]{R. Brock},
\author[bou]{J.M. Butler},
\author[pan]{S. Chopra},
\author[haw]{M. Cummings},
\author[fna]{D. Denisov},
\author[fna]{H.T. Diehl},
\author[tat]{S.R. Dugad},
\author[msu]{P. Duggan},
\author[msu]{S. Fahey},
\author[fna]{E. Flattum},
\author[niu]{M. Fortner},
\author[roc]{V. Glebov},
\author[cin]{J. L. Gonz\'alez Sol\'{\i}s},
\author[fna]{D.R. Green},
\author[niu]{J. Green},
\author[fna]{N. Grossman},
\author[tat]{A. Gupta},
\author[fna]{H. Haggerty},
\author[fna]{S. Hansen},
\author[msu]{R. Hatcher},
\author[niu]{D. Hedin},
\author[cin]{R. Hern\'andez-Montoya},
\author[fna]{T. Hu},
\author[fna]{S. Igarashi}
\author[fna]{A.S. Ito},
\author[msu]{S.A. Jerger},
\author[arz]{K. Johns},
\author[tat]{S.D. Kalmani},
\author[pan]{J.M. Kohli},
\author[tat]{M.R. Krishnaswamy},
\author[niu]{R. Markeloff},
\author[ind]{T. Marshall},
\author[lan]{T. McMahon},
\author[fna]{D. Miller},
\author[tat]{N.K. Mondal},
\author[tat]{P. Nagaraj},
\author[tat]{V.S. Narasimham},
\author[arz]{A. Narayanan},
\author[cbp]{M. Nicola},
\author[msu]{M. Nila},
\author[tat]{N. Parua},
\author[msu]{B.G. Pope},
\author[fna]{P.Z. Quintas},
\author[tat]{M.V.S. Rao},
\author[tat]{L.V. Reddy},
\author[fna]{T. Regan},
\author[msu]{T. Rockwell},
\author[tat]{B. Satyanarayana},
\author[tat]{H.C. Shankar},
\author[pan]{J.B. Singh},
\author[pan]{P.M. Sood},
\author[fna]{A. Taketani},
\author[tat]{P.R. Vishwanath},
\author[nor]{J. Wilcox},
\author[nor]{D.R. Wood},
\author[fna]{R. Yamada},
\author[nor]{T. Yasuda}

\address[arz] {University of Arizona, Tucson, Arizona 85721, USA}
\address[bou] {Boston University, Boston, Massachusetts 02215, USA}
\address[cbp] {LAFEX, Centro Brasileiro de Pesquisas F{\'\i}sicas, Rio de Janeiro, Brazil}
\address[cin] {CINVESTAV, Mexico City, Mexico}
\address[del] {Delhi University, Delhi, India 110007}
\address[fna] {Fermi National Accelerator Laboratory, Batavia, Illinois 60510, USA}
\address[haw] {University of Hawaii, Honolulu, Hawaii 96822, USA}
\address[ind] {Indiana University, Bloomington, Indiana 47405, USA}
\address[lan] {Langston University, Langston, Oklahoma 73050, USA}
\address[msu] {Michigan State University, East Lansing, Michigan 48824, USA}
\address[nor] {Northeastern University, Boston, Massachusetts 02115, USA}
\address[niu] {Northern Illinois University, DeKalb, Illinois 60115, USA}
\address[pan] {University of Panjab, Chandigarh 16-00-14, India}
\address[roc] {University of Rochester, Rochester, New York 14627, USA}
\address[tat] {Tata Institute of Fundamental Research, Colaba, Mumbai 400005, India}

\vspace{0.1in}
(for the D\O\ Collaboration)
\begin{abstract}
We present the results of an upgrade to the D\O\ muon system.  Scintillating
counters have been added to the existing central
D\O\ muon system to provide rejection
for cosmic ray muons and out-of-time background, and to provide additional fast
timing information for muons in an upgraded Tevatron.
Performance and results from the 1994--1996 Tevatron run are presented.
\end{abstract}
\end{frontmatter}
Submitted to Nuclear Instruments and Methods in Physics Research, \\
Section A

\newpage
\setcounter{page}{1}
\setcounter{footnote}{0}
\pagestyle{plain}

\section{Introduction}
\begin{figure}[h]
\vskip 2.0cm
 \begin{center}
  \begin{tabular}{c}
    \epsfxsize = 8cm \epsffile[100 150 550 550]{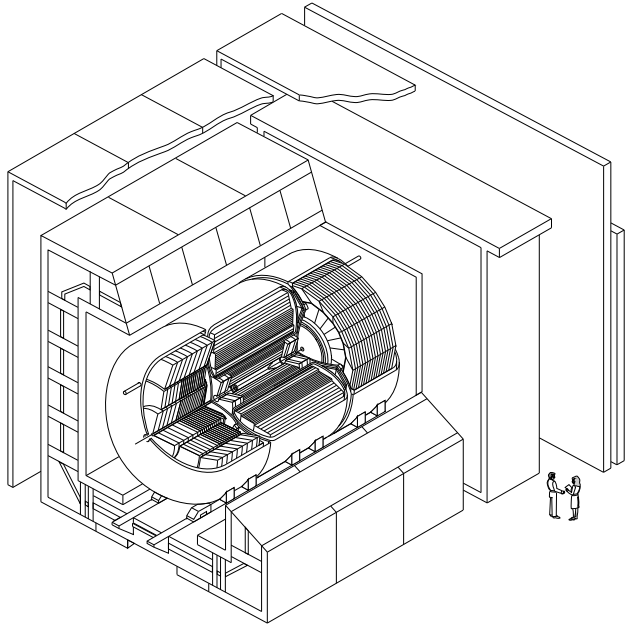}
  \end{tabular}
 \end{center}
\caption{An isometric view of the D\O\ detector.}
\label{fig:det_iso}
\end{figure}

The D\O\ detector at the Fermilab Tevatron was completed in 1992 and
took data during the 1992-1993 Tevatron Collider run
(Run~1a) and from 1994-1996 (Run~1b and 1c).
This paper describes a scintillator
upgrade to the D\O\ wide angle muon system which was fabricated for
Run~1b and 1c.  A cutaway isometric view of the detector is shown in
Fig.~\ref{fig:det_iso} where
the central detectors, calorimeter, and muon detectors are shown.

The D\O\ wide angle muon system consists of three separate solid iron
toroid magnets with three sets of proportional drift tube
chambers (PDT's) to measure track coordinates for each
muon.  The first set of chambers (A layer) are located
between the beamline and the toroid and
provide a closely spaced set of measurements of the muon track and
determine the entrance point into the toroid.
The second (B layer) and third (C layer) set of chambers, located
after the toroid, are separated by 1 m and yield the exit direction.
The bend in the toroid is in
a plane which approximately goes through the center of the Tevatron
beam pipe.

Due to the large PDT drift space the drift times
extend to as much as 0.9 $\mu$sec.  The accelerator operated with
6 bunches in the collider for Run~1 which gave
a 3.5 $\mu$sec time gap between collisions.  The next collider run (Run~2)
is scheduled to operate with 36 or more bunches which will have a
crossing time of 0.396 or 0.132~$\mu$sec.  In order to label the
muon in the PDT with the appropriate bunch crossing, a fast timing
reference will be required. Furthermore, the long livetime coupled
with an increased instantaneous luminosity would increase the
cosmic ray coincidence rate to an unacceptable level.  The solution
to this problem, for the central region, was to add scintillation
counters to the outside of the muon system.  The counters were
installed in Run~1b as part
of the upgrade for Run~2.  The counters provided cosmic
ray rejection.  In addition the counters provided rejection
of backscattered particles which were found to have a
significant effect on the trigger.

The scintillation counters were installed on the
sides and top of the central muon system (CF).
The scintillation counters
spanned a region of $\mid\!\eta\!\mid\leq 1$ and
covered 6 of
the 8 octants in azimuth where each octant consisted of 40
scintillation counters.
The octants were numbered from 0 to 7 in increasing azimuth,
such that octants 1 and 2 covered the top quadrant of the detector.
Counters were not installed on the bottom or ends
($\mid\!\eta\!\mid\geq 1$) of the D\O\ detector.
Bottom counters, which could not be installed while the detector was in the
collision hall, will be added before Run~2.
End counters required a different design due
to the high multiplicities in this
region.  End counters will also be added for Run~2. Prototype
end counters were tested during Run~1b and 1c.
This document describes only the Run~1 counters.

\begin{figure}[t]
 \begin{center}
    \begin{tabular}{c}
  \epsfxsize = 8cm \epsffile[100 150 550 550]{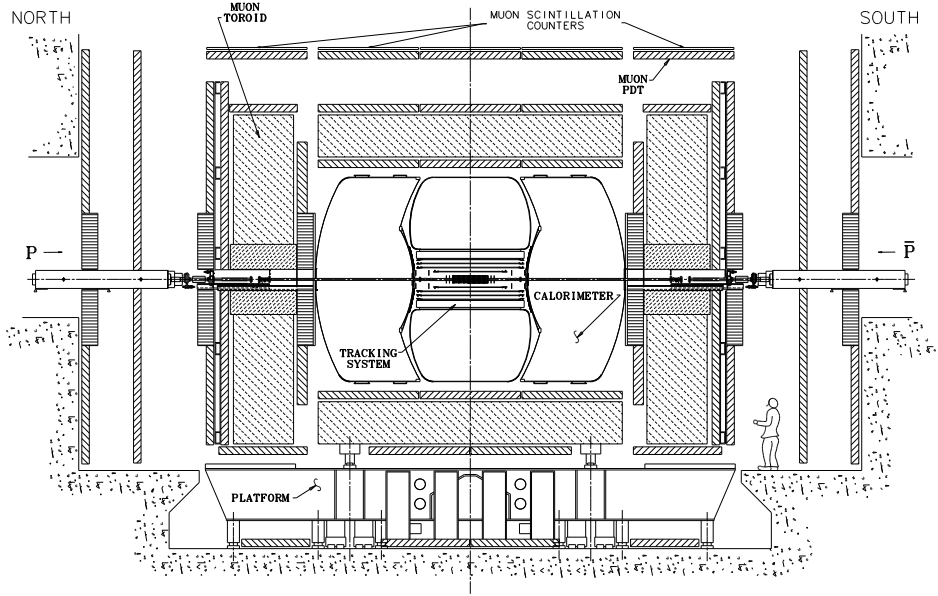}
    \end{tabular}
 \end{center}
\vskip 2.0cm
\caption{An elevation view of the D\O\ detector showing the location
of the muon scintillation counters.}
\label{fig:det_elev}
\end{figure}

Figure~\ref{fig:det_elev} is an elevation view of the detector, which shows the
detector
systems mentioned above and the location of the top
scintillation counters. Also shown is the Tevatron beam pipe
centered within the detector.  The muon system has been described in previous
publications~\cite{det:nim} and
so will not be discussed in any more detail here.

\section{Scintillation Counter Design}
\begin{figure}[htb]
 \begin{center}
  \begin{tabular}{c}
    \epsfxsize = 10cm \epsffile[100 150 550 550]{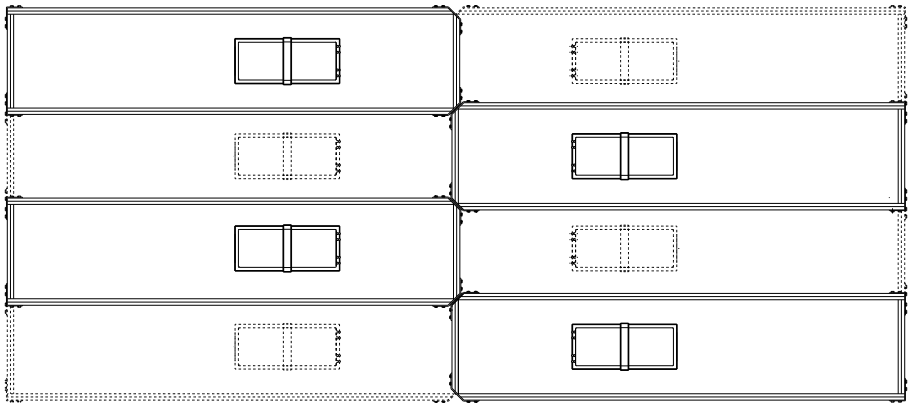}
  \end{tabular}
 \end{center}
\vskip -3.0 cm
\caption{The nested configuration of eight scintillation counters for a muon
PDT module. The counters represented by the solid lines are offset in
height relative to the counters drawn with the dotted lines.}
\label{fig:scint_pdt}
\end{figure}

\begin{figure}[htb]
 \begin{center}
  \begin{tabular}{c}
    \epsfxsize = 10cm \epsffile[100 150 550 550]{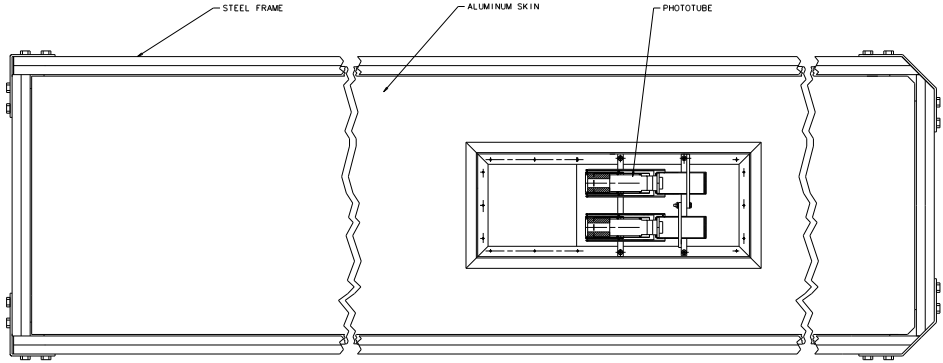}
  \end{tabular}
 \end{center}
\vskip -4.0 cm
\caption{The top view of a muon scintillation
counter showing the support frame and the location
of the photomultiplier tubes and cover.}
\label{fig:scint_counter_top}
\end{figure}

The 240 scintillation counters cover the outside layer of PDTs in the
central region.  There are three different lengths of
scintillation counters
to match the three different sizes of the
PDT's, with the long dimension of the scintillation counter chosen to
match the long dimension of the PDT cell (see Table~\ref{tab:scint_size}).
There are 8 counters
per PDT module in two overlapping layers as shown in
Fig.~\ref{fig:scint_pdt}.
The corners on one end of the scintillator are cut at
a 45$^\circ$ angle to allow counters to be nested.  The final
configuration of two layers provided approximately 1 to 2~cm overlap
to minimize the dead area (see Fig.~\ref{fig:scint_pdt}).
Figure~\ref{fig:scint_counter_top} shows a close up view of
the top of a scintillation counter.

\begin{figure}[h]
 \begin{center}
  \begin{tabular}{l}
    \epsfxsize = 10cm \epsffile[100 150 550 550]{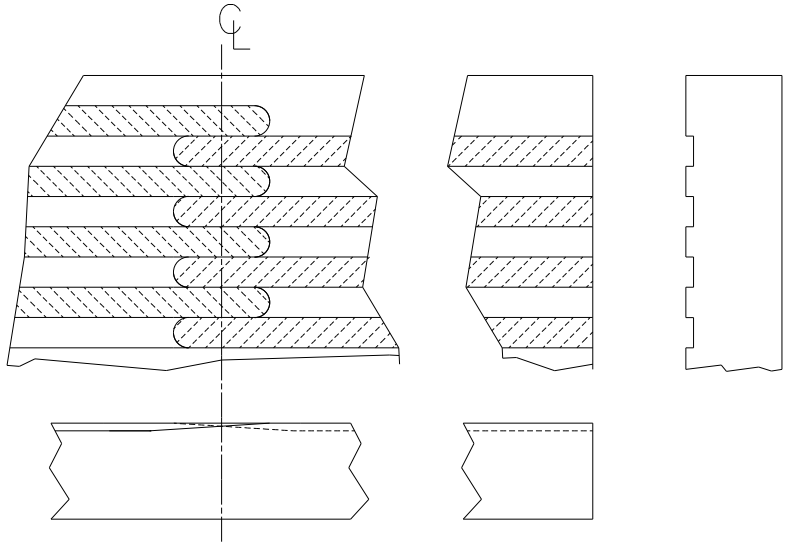}
  \end{tabular}
 \end{center}
\vskip -0.5 cm
\caption{The top, side, and end views of the grooves in the scintillator are
shown.  In each groove are epoxied four waveshifting fibers.}
\label{fig:sc_grv}
\end{figure}

The scintillation counters were
fabricated at Fermilab and the Tata Institute of
Fundamental Research (TIFR)
in India using Bicron 404A scintillator~\cite{bic:404a}.
\begin{table}
    \begin{center}
        \begin{tabular}{ccccc}
Length (in) & Width (in) & Depth (in) & grooves & fibers/pmt \\ \hline
 81.5  & 25.0 & 0.5 & 78/79 & 314\\
108.0  & 25.0 & 0.5 & 78/79 & 314\\
113.0  & 25.0 & 0.5 & 78/79 & 314\\ \hline
        \end{tabular}
    \end{center}
\caption{The three sizes of muon scintillation counters.}
\label{tab:scint_size}
\end{table}
The Bicron 404A scintillator was
chosen because of its low cost and good performance. It was
approximately a factor of two less expensive
than comparable scintillators.
The relatively short attenuation length of this scintillator was not a
problem because the light was
collected within a short distance by wavelength shifting fibers.
The scintillator was machined with 4~mm wide straight
grooves 1.75~mm deep on one side (see Fig.~\ref{fig:sc_grv}).
The grooves on each half of the scintillator were offset so that
they would not overlap in the center.  The half of the scintillator with 
the 45$^\circ$ corners had one more groove than the other half.  
The machining was done using computer controlled milling heads
at Lab 8 at Fermilab, a commercial company in Rockford IL~\cite{rockford},
and at the central workshop of Bhabha Atomic Research Center,
Bombay~\cite{bhabha}.
The
grooves were machined with 8~mm spacing and covered roughly half the area
of one face of the scintillator.
Four 1~mm diameter Bicron BCF91A wavelength shifting
fibers~\cite{bic:91a} were placed in each groove.
The fibers were brought to the center of the
counter to be viewed by phototubes.  Each fiber covers half the length of the
counter. This center readout geometry was chosen to reduce the attenuation of
the light in the fiber and also reduce the time jitter due to light transmission
in the fiber.  For the 81.5'' counters built at TIFR, the edge grooves
were loaded with
Kuraray Y-11~\cite{kuraray} double clad fibers to increase the light output
at the edge of the counters.  The double clad fibers provide
a factor of 1.8 more light than a single clad fiber due to
a more efficient capture of the wavelength shifted light and also
due to a longer attenuation length in the fiber.  The double clad fibers
were only used on the edges
since they were about five times more expensive than single clad fiber.

The fibers were held in place with Bicron 600 optical
epoxy~\cite{bic:600} at seven
locations along the groove.  Five minute
epoxy~\cite{5min:epoxy} was used at the ends of counters
where the fibers extended beyond the scintillator.  Both ends of the
scintillator were diamond polished using a diamond cutter manufactured at
Michigan State University (MSU)~\cite{MSU:cut}
and brought to Fermilab and retrofitted
for this project.  For the counters made at TIFR, a custom
made diamond polisher machine~\cite{bombay_co} was constructed.
In order to increase the light yield a 1/32'' anodized
aluminum sheet~\cite{ever:brite} was placed on the diamond polished ends and
held in place with aluminized mylar tape. The fibers going to the 45$^\circ$ cut
corners were sputtered with aluminum at Fermilab's Lab 7 facility,
and with an electron beam heating method at TIFR.

Two of the four fibers
from each groove were epoxied into the hole of an acrylic
cookie using five minute epoxy.  The other two
fibers from each groove were bundled and bonded to a second cookie.
The fibers were rough cut with a saw and then the ends of
cookies and the fibers were polished with the diamond cutter.

The counters were then wrapped with a Tyvek~\cite{dup:tyvek,dup:white}
sheet with a hole cut for the fibers and cookies.  Tests comparing the Tyvek
wrap to aluminum foil and aluminized mylar showed that the Tyvek improved the
counter response by about 10\%.
A 3'' wide black tape was used to secure the Tyvek sheet and
cover any holes or seams.
A 1/4'' thick, closed cell polyethylene foam was placed on the
top of the scintillator.
The top and bottom surface
of the scintillator were then
covered with a 0.020'' thick aluminum
sheet (see Fig.~\ref{fig:scint_counter_top}).
An 8''$\times$7 13/32''
opening was cut in the top
sheet for the fiber bundles
and an aluminum angle lip was mounted around this hole.
The aluminum sheets were taped together around the perimeter of the counter.
In order to provide the support for the
counter, commercially available steel channel
frame~\cite{unist:clone} was placed over the
aluminum sheets around the edges and bolted together.
The counters then had metal bands attached at the center
for additional support.

\begin{figure}[h]
 \begin{center}
  \begin{tabular}{l}
    \epsfxsize = 10cm \epsffile[100 150 550 550]{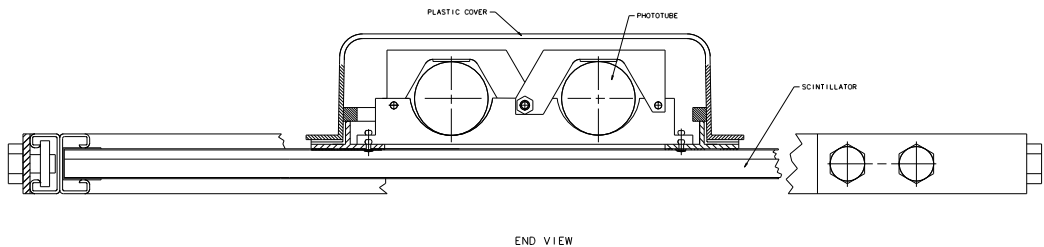}
  \end{tabular}
 \end{center}
\vskip -4.0 cm
\caption{The end view of a muon scintillation counter.}
\label{fig:scint_counter_end}
\end{figure}

A plastic cover was used to provide access to the photomultiplier tubes.
The plastic cover was fastened to the aluminum angle with velcro and the seams
were taped.
SHV and BNC connectors for the
high voltage and output signal were bolted to the cover
for each phototube.
Figure~\ref{fig:scint_counter_end} shows an end view of a
scintillation counter detailing the location of the photomultiplier tubes and
cover.
    \subsection{Fiber}
\label{sec:fiber}
The properties of the Bicron 91A waveshifting fiber were
studied as part of the counter research and development
work.  The Bicron 91A waveshifting fiber
was designed to be used in conjunction with the Bicron 404A scintillator.
The light output from the fiber was tested with a variety
of reflective surfaces at the far end of the fiber.  The fibers were tested with
the far end having a razor cut alone, diamond polished alone, diamond polished with
aluminized mylar tape added, and diamond polished with the end sputtered with aluminum.
The sputtered aluminum had the highest light output followed by the
aluminized tape.

The measurements were made at the fiber testing facility at
Fermilab~\cite{fiber_test}. The
1~mm diameter fibers were cut into 3~m lengths.
Excitation was done at 30~cm intervals throughout the length of the fiber using
ultraviolet light from a pencil source~\cite{oriel:6035}.  The light from the fiber
was detected using a silicon photodiode~\cite{optron:phdiode} whose current was measured
with a picoammeter.  The entire system was operated under microcomputer control.

Each end of the fiber was epoxied into an acrylic ferrule and one end
of the fiber was coated with optical grease~\cite{bic:grease} and
clamped, using the ferrule, to the photodiode.

The measured intensity was fit by the sum of two exponentials of the
form:
\begin{eqnarray}
    {\rm I}(x) &=&{\rm A}_{1}\cdot exp(-x/\alpha_{1}) + {\rm A}_{2}\cdot exp(-x/\alpha_{2})
\end{eqnarray}
where A1, A2, $\alpha_{1}$, and $\alpha_{2}$ are the fitted parameters and
$x$ is the position along the fiber.
\begin{figure}[h]
 \begin{center}
  \begin{tabular}{c}
    \epsfxsize = 8cm \epsffile{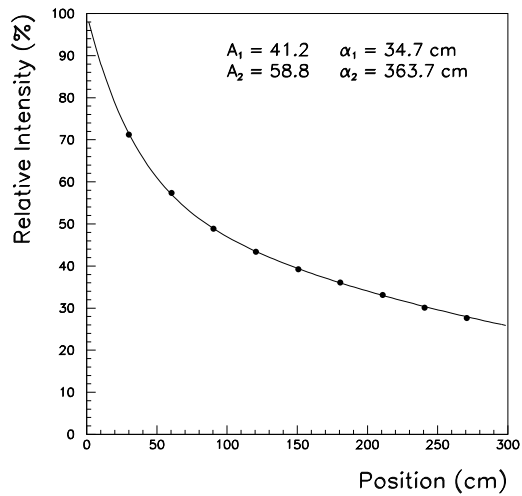}
  \end{tabular}
 \end{center}
\caption{ The relative intensity as a function of the excitation location
along the fiber.}
\label{fig:int_fiber_fit}
\end{figure}
Figure~\ref{fig:int_fiber_fit} shows the
fit to the relative intensity for an arbitrary fiber.  The long attenuation
term is due to the light in the core of the fiber and the short attenuation
length term is due to the light transmitted in the cladding of the fiber.

The maximum light output was observed for diamond polished and
sputtered fibers which yielded 65\% more light output than the razor
cut fibers.  The light output with diamond polished fibers and aluminized mylar
tape was about 10\% less than the aluminum sputtered fiber.

The above studies led us initially to use the diamond
polished fibers with an aluminized mylar tape over the sputtered fibers,
because the time required to sputter the 150,000 fibers
exceeded the time limit for construction of the system.
However, for the final design, an anodized aluminum strip~\cite{ever:brite}
was placed over the ends since
the aluminized tape formed bubbles in the adhesive.
The light output from this design
was shown to be stable over time.
    \subsection{Phototube}
The 1.5'' diameter EMI 9902KA photomultiplier tube~\cite{ven:EMI}
was selected
for readout of the light produced by the scintillator.
A number of criteria determined the selection of
this phototube.
Space limitations dictated the diameter,
the phototube had to match the light
emission characteristics of the waveshifter fiber, thus the photocathode
required a spectral response which extended into the green region.
Also required were an
amplification of $10^{7}$ and  a dark current of less than 10 nA.
The tubes were required to be fast and have a rise time of less than
3 nsec.

\begin{figure}[th]
\vskip -1.cm
 \begin{center}
  \begin{tabular}{c}
    \epsfxsize = 9cm \epsffile[100 150 550 550]{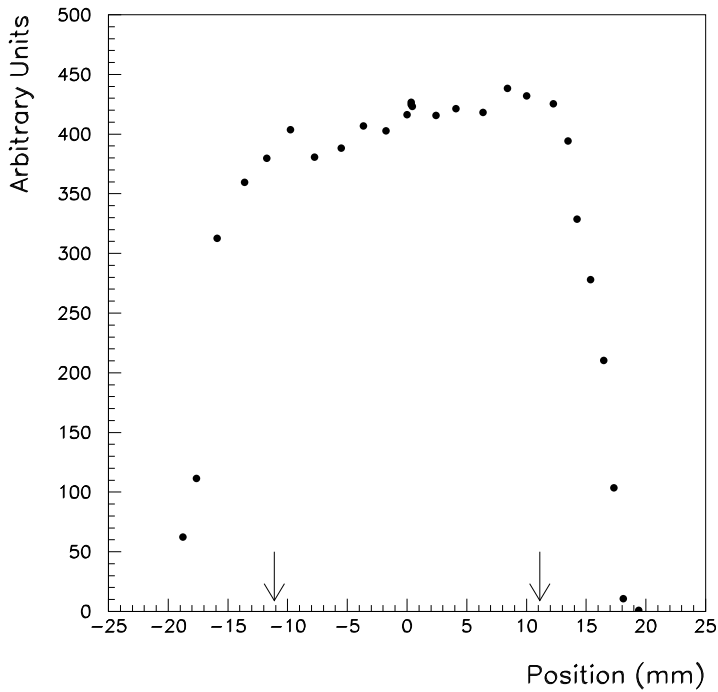}
  \end{tabular}
 \end{center}
\vskip 2. cm
\caption{Relative response across the face of the photocathode for
an EMI 9902KA phototube where zero is the center of the photocathode.
The arrows delineate the region where the fiber is
incident upon the photocathode.}
\label{fig:pmt_face}
\end{figure}

Since fibers from different parts of the counter were randomly
positioned on the photocathode face, we investigated the
response of the EMI 9902KA as a function of position on the
photocathode at Fermilab's phototube testing facility.  Light 
from an ultraviolet laser was brought
via an optical fiber to illuminate $\sim$2 mm spot
on the PMT photocathode.  The response, in arbitrary units, as a
function of position is shown in Fig.~\ref{fig:pmt_face}
for a typical 9902KA phototube.
The fibers from our counters formed a bundle 7/8" in diameter at the
photocathode and so were well matched to the high response region
of the PMT.

\begin{figure}[th]
\vskip 1.0 cm
 \begin{center}
  \begin{tabular}{c}
    \epsfxsize = 9cm \epsffile[100 150 550 550]{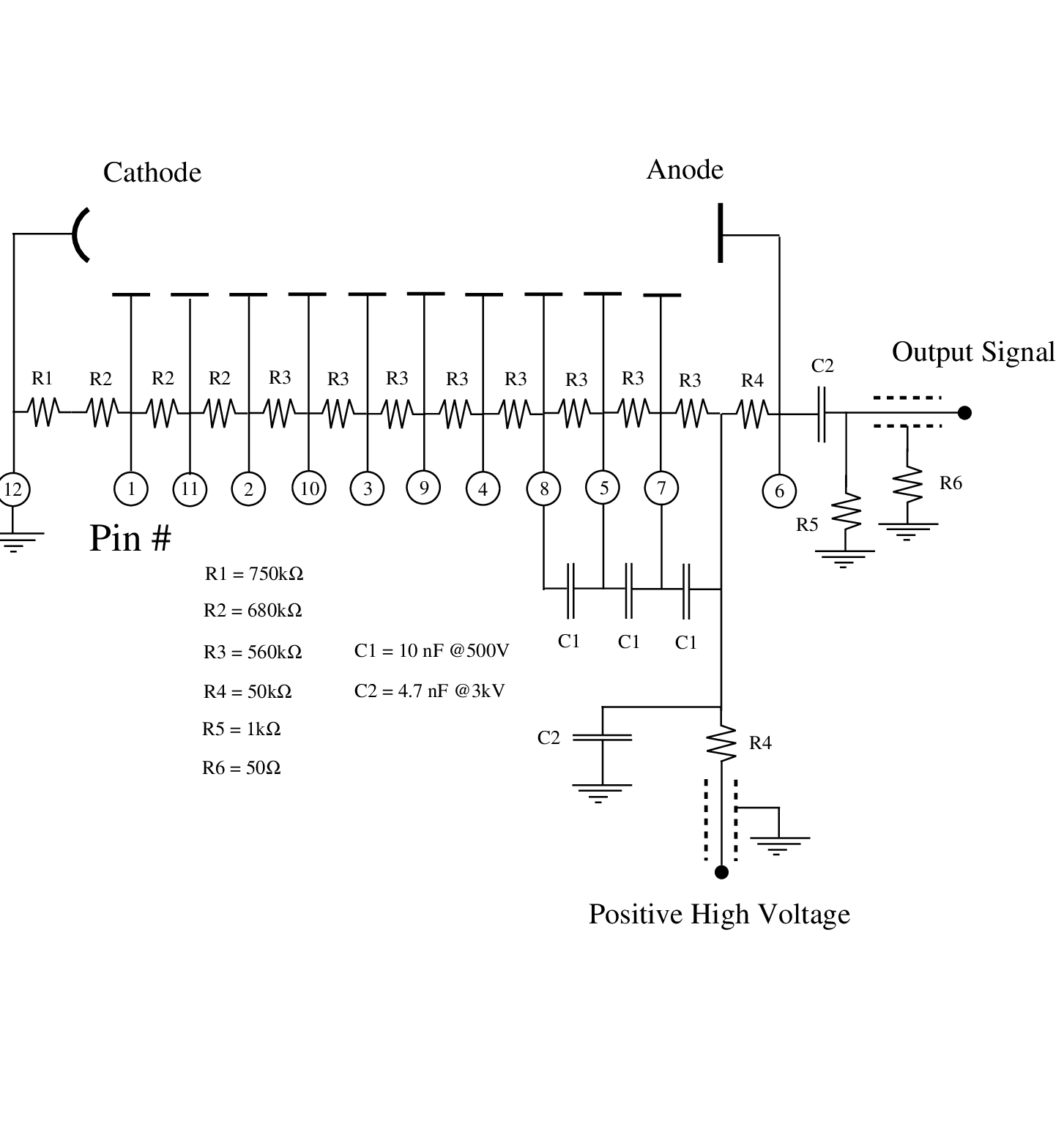}
  \end{tabular}
 \end{center}
\caption{The circuit diagram for the photomultiplier base.}
\label{fig:pmt_circuit}
\end{figure}

For compatibility with the D\O\ High Voltage system, the photomultiplier
base voltage divider and readout circuit (see Fig.~\ref{fig:pmt_circuit})
was required to draw a
low current: 0.2~mA at a cathode-anode voltage of +2~kV.
The resistor values chosen
to  meet this requirement were 2.5:1.2:1.2:1.0
for the first, second,  third, and remaining stages.
To minimize output noise into the scintillator electronics
readout cards the output signal was filtered to reduce low frequency(ac) noise
and the signal ground
isolated from the common (HV) ground. The input HV was also filtered and
capacitors were used across the last three stages to prevent voltage sagging.

Due to tight space constraints in several areas of the detector where
counters were to be installed, the base assembly was required to be
compact. The  base circuit was mounted on a 1.75''$\times$1.5''
printed circuit board (PCB), connected by a
standoff to a circular  PCB 1.5'' in diameter which was fitted onto a Hamamatsu
socket~\cite{ven:Hama}. This assembly was conformally
coated and fitted inside a 2''$\times$4''
cylindrical aluminum container, which
provided electromagnetic shielding, with the socket protruding through
one end. Readout and HV connections were made through an isolated
BNC and an SHV bulkhead mounted on a 2'' diameter aluminum cap which
was secured onto the end of the cylinder. To insure the
consistency of
the large number (800) of bases produced, the PCBs were
manufactured~\cite{ven:pcb}, stuffed, and
assembled by outside vendors~\cite{ven:stuff}.

The phototube and base were secured to the cookie using a spring mechanism
which attached to the cookie and to the aluminum cap.  This ensured the
photocathode was centered on the cookie and flush with the surface.
Surrounding this assembly was placed a Model 17P37
magnetic shield~\cite{mag_shield}
to protect the phototube from magnetic fields.
In addition, the entire assembly was surrounded by a
second 1/8'' thick, iron shield to ensure good magnetic field protection.
\section{Scintillation Counter Electronics}
\begin{figure}[th]
\vskip 1.0 cm
 \begin{center}
  \begin{tabular}{c}
    \epsfxsize =15cm \epsffile{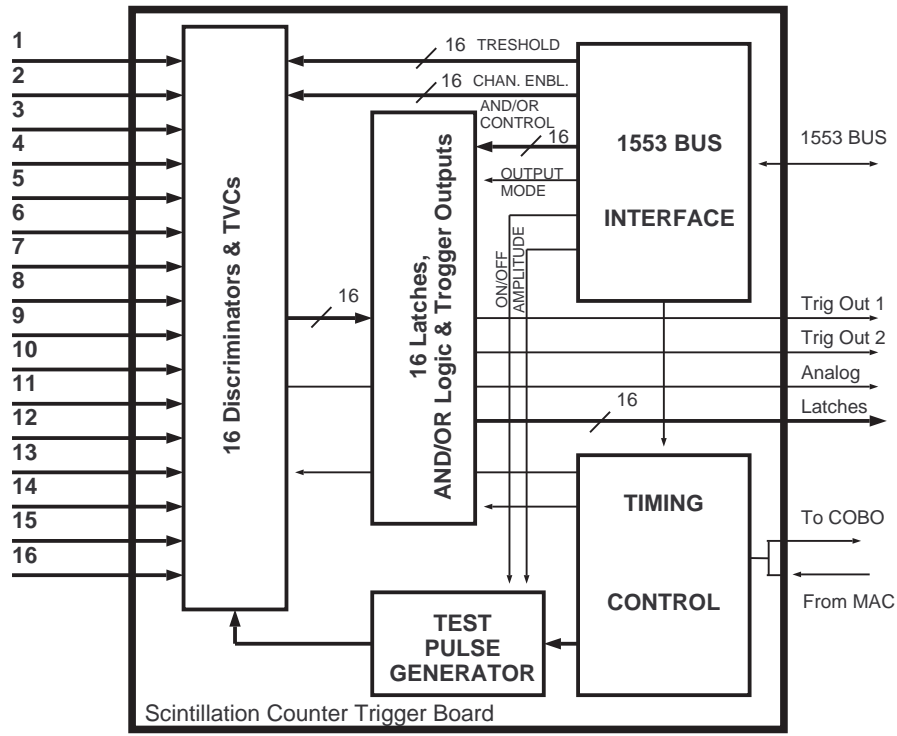}
  \end{tabular}
 \end{center}
\caption{The block diagram for the scintillation counter trigger board.}
\label{fig:sctb_1}
\end{figure}

The block diagram for the scintillation counter trigger board (SCTB) is
shown in Fig.~\ref{fig:sctb_1}.
There was one electronics board for each group of 8
scintillation counters (16 phototubes) which were mounted on a muon system
PDT.  Each board had 16 signal processing
channels, trigger logic, a test pulse generator, timing controls and
a 1553B bus interface for control and status.  Each scintillator channel
could be remotely enabled or disabled using an enable bit register. The
1553 bus is the standard used by the D\O\ experiment for downloading and
monitoring the front end electronics.  The timing control circuit was a
programmable gate generator linked to the transition of a muon system
timing signal. The trigger logic generated output trigger signals based on
the inputs and the AND/OR settings for each channel.  The test pulse
generator simulated the scintillation counters signal and could be
synchronized to different sources.

\begin{figure}[th]
\vskip -5.0 cm
 \begin{center}
  \begin{tabular}{c}
    \epsfxsize = 14cm \epsffile{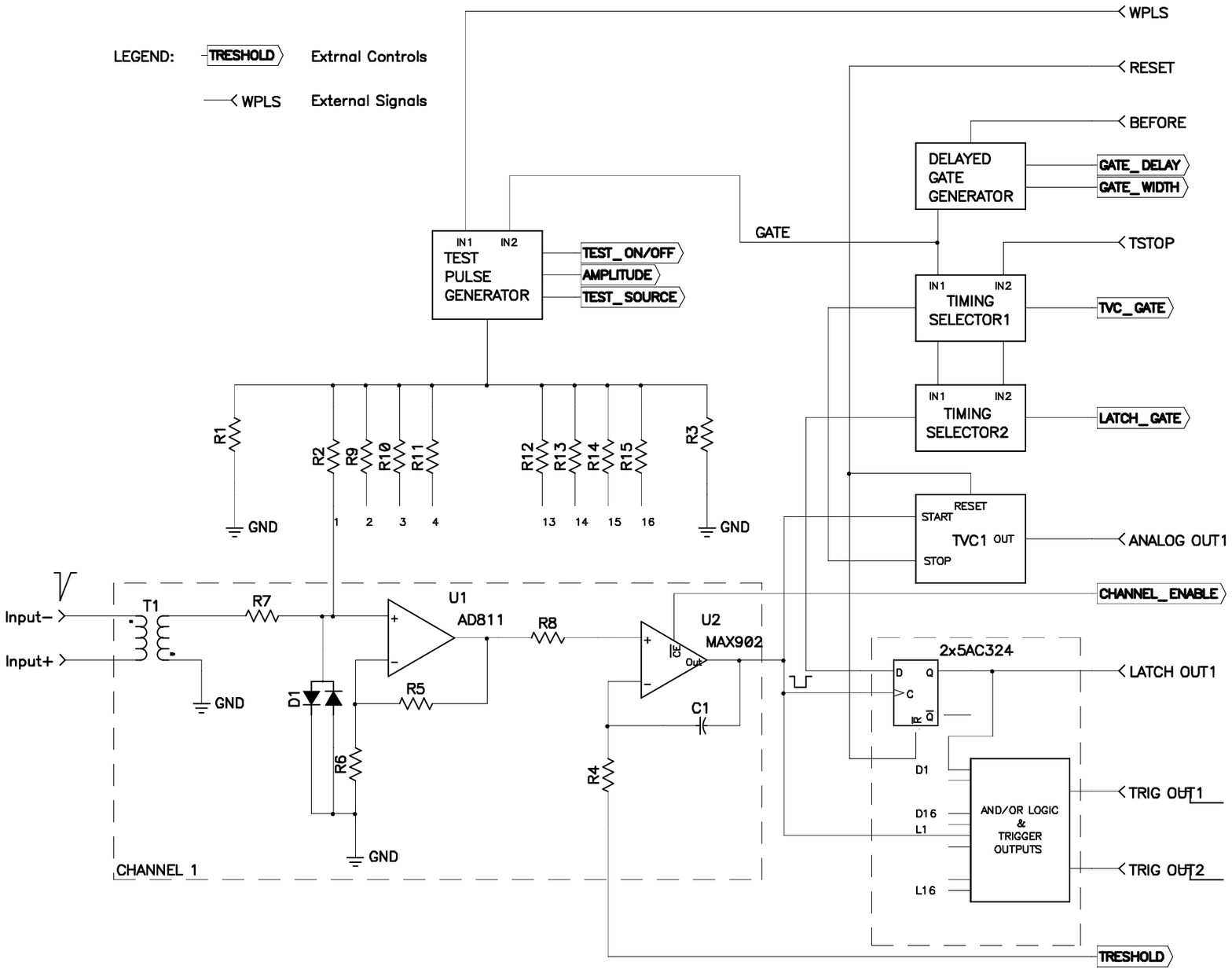}
  \end{tabular}
 \end{center}
\vskip -3cm
\caption{The diagram of the amplifier/discriminator, trigger latch and TVC for a
SCTB channel.}
\label{fig:sctb_2}
\end{figure}

Each SCTB channel had an amplifier/discriminator, trigger latch and
time-to-voltage converter (TVC) (see Fig.~\ref{fig:sctb_2}).
DC isolation of the SCTB
inputs from the common HV ground connected to each phototube base was
provided.  Each input was transformer coupled and has optional jumpers to
isolate primary windings from the SCTB ground plane.   The
threshold level for each channel could be independently and remotely set from
10~mV to 255~mV.  Each channel was supplied with a common test signal from
the test pulse generator.  The amplitude of the signal was digitally controlled
which allowed for remote testing of the particular channel during the run.

Each SCTB channel had a TVC connected to its
discriminator output.  There were two TVC ranges available, a 500 nsec
range for normal data taking and a 2~$\mu$sec range for test purposes.  The
internal gate width and delay for the 16 channels could be remotely
varied up to 510~nsec in 2~nsec steps.

Each of the two trigger outputs produced a 24 mA output signal which exceeds
the standard 16 mA NIM requirement.  Each pair of the 16 channel latch
outputs could be ANDed or ORed depending on the selection of the
corresponding bit in the AND/OR selection byte.  Eight product signals were
then ORed together producing the trigger signal output.  Each trigger output
had its own AND/OR control byte which can be set separately.  The trigger
signals from each scintillator board were brought to the NIM logic
system described in section~\ref{sec:veto}.

The scintillator data, consisting of digitized TVC output
and latch bits(see Fig.~\ref{fig:sctb_1}) were recorded using the muon
electronics system~\cite{det:nim}.  This system treated the scintillator counter
information as an additional layer of muon PDT for the purpose of data
recording.
\section{Production}
    \subsection{Scintillator Testing}
One scintillator from each batch 
was tested prior to
having the grooves machined into the surface.  The test was performed by placing
the scintillator in a light tight box with a 24''$\times$24'' 
anodized aluminum sheet
placed underneath the scintillator~\cite{ever:brite}.  Using a telescope of
three scintillator paddles, a cosmic ray trigger was generated and the light
from the scintillator was measured using a 12''$\times$12'' 
sheet of Bicron 91A fiber
glued together to form a flat surface using optical 
epoxy~\cite{bic:600} and coupled to an EMI 9902KA phototube.  The signal area
was digitized using a LeCroy QVT and was compared to the signal from
a reference sheet of scintillator.  The batches tested varied 
in measured pulse height by less than 6\% and no scintillator was rejected by
these tests.  
    \subsection{Fiber Testing}
Every 500~m reel of fiber used in the scintillation counter production was
tested.   The fiber used in the counters were required to pass a light
attenuation and an absolute light output test.
In all, 700 samples of fibers were tested

From the test of each 3~m fiber the following quantities were recorded
\begin{eqnarray}
{\rm I}_3 &= &
    {\rm Intensity\ of\ light\ for\ a\ source\ at\ } 30 \:{\rm cm} \\
{\rm I}_{15} &= &
    {\rm Intensity\ of\ light\ for\ a\ source\ at\ } 150 \:{\rm cm} \\
{\rm I}_{27} &= &
    {\rm Intensity\ of\ light\ for\ a\ source\ at\ } 270 \:{\rm cm}
\end{eqnarray}
where the equipment used to test the fiber is the same as discussed in
section~\ref{sec:fiber}.
Similarly, the three intensities were recorded for special fibers chosen to
be used as a reference: R$_{3}$, R$_{15}$, and R$_{27}$.
The reference fibers were needed to
normalize the day to day variations in the fiber testing apparatus.  For
normalization purposes the following ratios were
calculated for each group of fiber measurements
\begin{eqnarray}
\label{eq:ir3}
    {\rm IR}_{3} &=& {\rm I}_{3}/{\rm R}_{3}  \\
\label{eq:ir15}
    {\rm IR}_{15} &=& {\rm I}_{15}/{\rm R}_{15}  \\
\label{eq:ir27}
    {\rm IR}_{27} &=& {\rm I}_{27}/{\rm R}_{27}.
\end{eqnarray}
The intensity variation due to the source and photodiode was tracked
over a period of 5 months.
The relative variation as a function of distance
changed during the 5 month testing period but
the variation in equations~\ref{eq:ir3},~\ref{eq:ir15},
and~\ref{eq:ir27} was minimal.

\begin{figure}[h]
 \begin{center}
  \begin{tabular}{cc}
    \epsfxsize = 7cm \epsffile{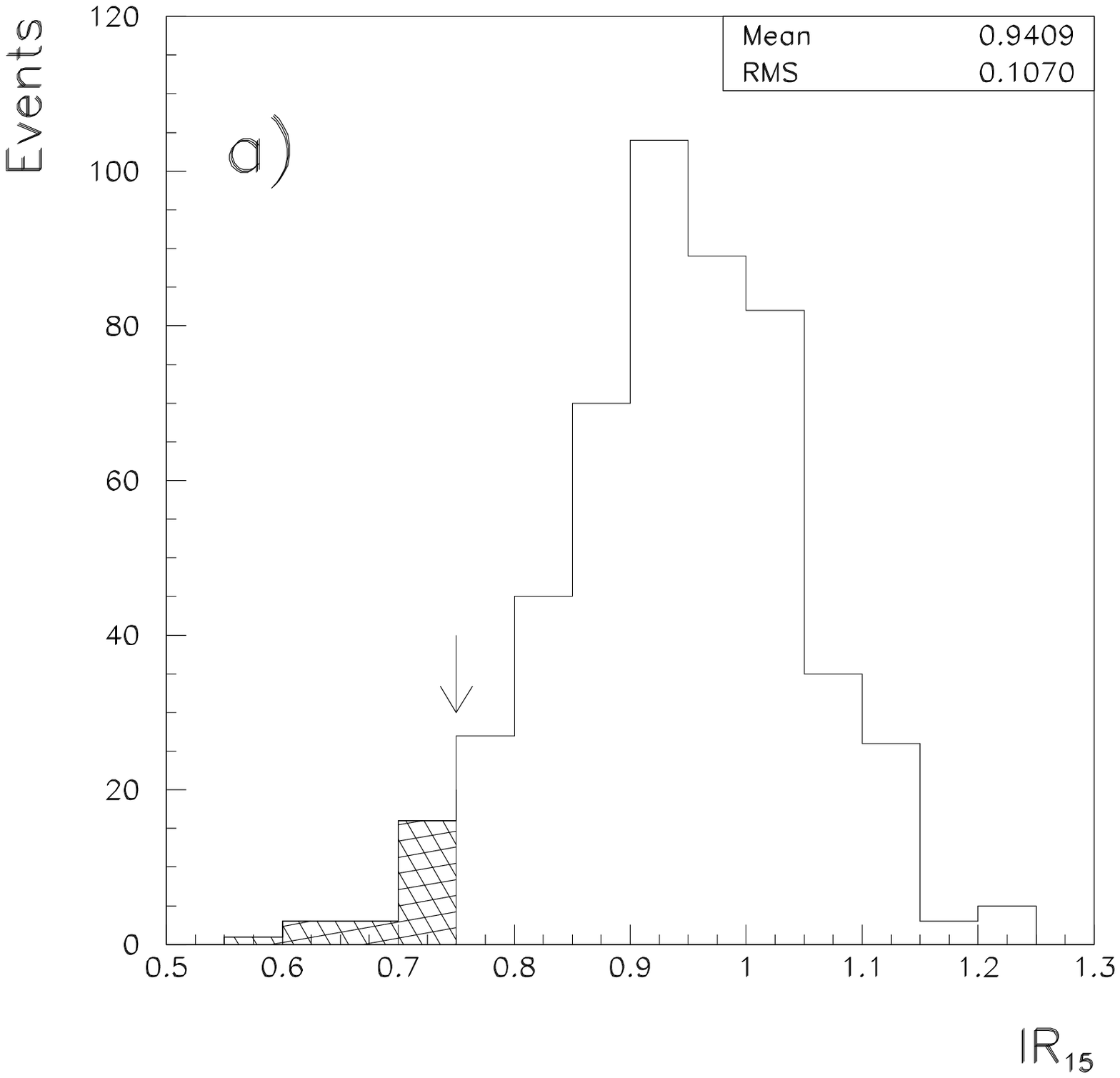} &
    \epsfxsize = 7cm \epsffile{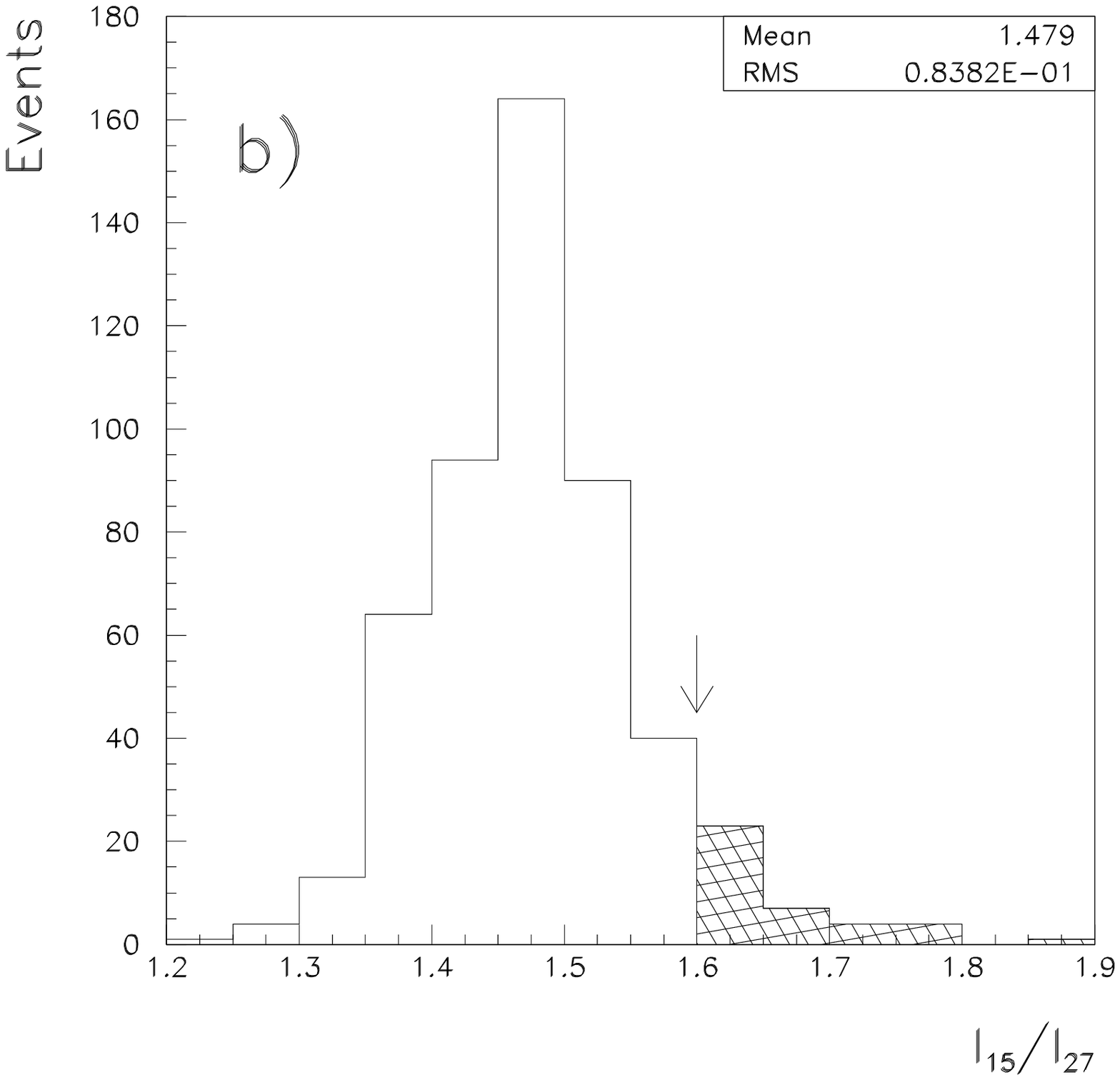} \\
  \end{tabular}
 \end{center}
\caption{(a)The distribution of the absolute light output parameter
IR$_{15}$. (b)The distribution of the attenuation length parameter
I$_{15}$/I$_{27}$. The hatched regions are the fibers that were 
removed by the cuts.}
\label{fig:fiber_atten}
\end{figure}

Deciding whether a particular fiber was accepted depended
upon the attenuation
length of the fiber and the absolute light output.
The attenuation length was
measured by calculating the ratio I$_{15}$/I$_{27}$ for each fiber.  By
making a cut of I$_{15}$/I$_{27}<1.6$, fibers with short
attenuation lengths were removed.   The results of this cut are shown in
Fig.~\ref{fig:fiber_atten}.
The absolute light output was checked by cutting on IR$_{15}$
for each fiber.  Fibers which had a value below
0.75 were rejected.  The results for this cut are shown
in Fig.~\ref{fig:fiber_atten}.
These cuts rejected about 9\% of the fiber tested.
    \subsection{Phototube Testing}
\label{sec:pmt}
The preliminary testing of the photomultiplier tubes and bases was done at
MSU.  The aim of this procedure was to check the
tubes and bases for proper operation, determine a rough level of dark
current noise, determine the voltage at which each tube-base
combination provided a $10^7$ gain, and match the tubes into groups of similar
operating voltage.

Tests of proper current, voltage division, and ground connections were made
on each base as a final production step by the assembly vendor.  In
addition, the tube-base output pulses were checked by oscilloscope as part of
the test process to verify correct pulse shape characteristics.  A
practical measure of tube noise was made by powering each tube at its
operating voltage in a dark box, applying a threshold of
30~mV to the tube output, and measuring the counting rate.
The 30~mV threshold was close to the threshold used on the detector readout
electronics.  

To determine the operating voltage of the tubes, a LED test assembly was
constructed.  A green (562~nm) LED was installed in an aluminum endpiece,
which was inserted in the end of the mu-metal shield and fixed in 
place. The LED was recessed and centered within the endpiece so that a
phototube inserted into the shield and butted up against the endpiece was a
constant distance from the light source.  This assembly was then mounted
inside the dark box.  A fast (8~nsec) voltage pulse with a 50~$\mu$sec
period was applied to the LED to provide an input light pulse.  The
charge of the output tube pulses was read out using a QVT multi-channel
analyzer with an internal threshold of 30~mV.  A Hamamatsu R580-17 
tube, previously measured for gain and output charge, was used as a reference 
to calibrate the system.  The operating voltage of the EMI tube was
defined as the voltage at which that tube gave an output pulse equal to that
of the Hamamatsu PMT.   Additionally, pulse height measurements were made at 
+50~V and $-$50~V around the operating voltage to provide information on
the gain variation.  Of the 800 tubes delivered 10 did not have any response
and about 50 had operating voltages above the maximum allowed value of 
1600~V.
These tubes were replaced by the vendor.  
    \subsection{Production Testing}
\label{sec:install_test}
%
%
A light leak test of the counters was done
after a counter was completed.  The test used a bank of flourescent
lights and scalars to check for leaks.
The counters were then placed in a light tight box which
allowed for the simultaneous test of three counters.
A Digital VAXstation Model
3100 was used to control the testing electronics and to
store the test results.
The workstation interfaced with the CAMAC equipment through a Jorway Model 111
PDP-11/CAMAC interface.
Since the objective of the production
test was to verify the counter performance, six reference phototubes and bases,
with an absolute gain within a 10\% range
were used. The counter testing was performed
using cosmic ray telescopes consisting
of three counters in coincidence (two above and one below the testing box)
at five locations (the middle of the counter and each of the four corners).
A trigger was formed by the logical OR of the five telescopes, and events
with more than one telescope hit were rejected offline.
A pattern unit was used to record which counters and telescopes were hit.
The information from the pattern unit was then used to determine the efficiency
of the counters (number of counter hits divided by number of triggers)
as a function of position.  CAMAC ADC's and TDC's were used
to measure the pulse height and timing for each tube.  When combined with
the pattern unit information, these values could be determined at each
telescope position.  CAMAC scalers were also used to monitor all the rates.
A completely automated procedure was developed
to run the test. Two runs, each lasting three hours,
were performed on each set of counters with the high voltage
set at the nominal value and nominal minus 200~V.  These two points allowed
a relative comparison to be made between all
counters.  The counters were all found to have efficiencies near 100\% and the
overall variation of the counters was small.  The only counter rejected
by the tests was one that was deliberately built with sub-standard fibers,
and consequently showed a reduced pulse height.

\begin{figure}[ht]
 \begin{center}
  \begin{tabular}{c}
    \epsfxsize = 12cm
        \epsffile[100 150 550 550]{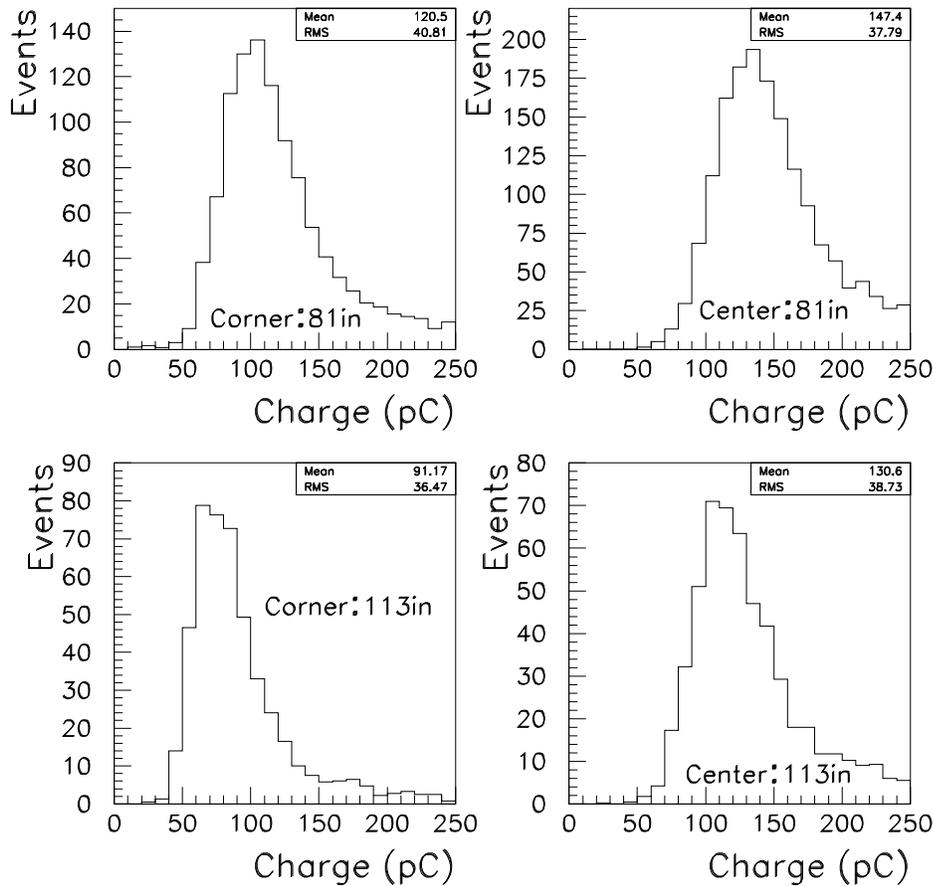}
  \end{tabular}
 \end{center}
\vskip 3.0cm
\caption{The pulse height distributions for the corner and center of the
81''(top) and 113''(bottom) counters.}
\label{fig:ave_pulse}
\end{figure}

\begin{figure}[ht]
 \begin{center}
  \begin{tabular}{c}
    \epsfxsize = 10cm
        \epsffile[100 150 550 550]{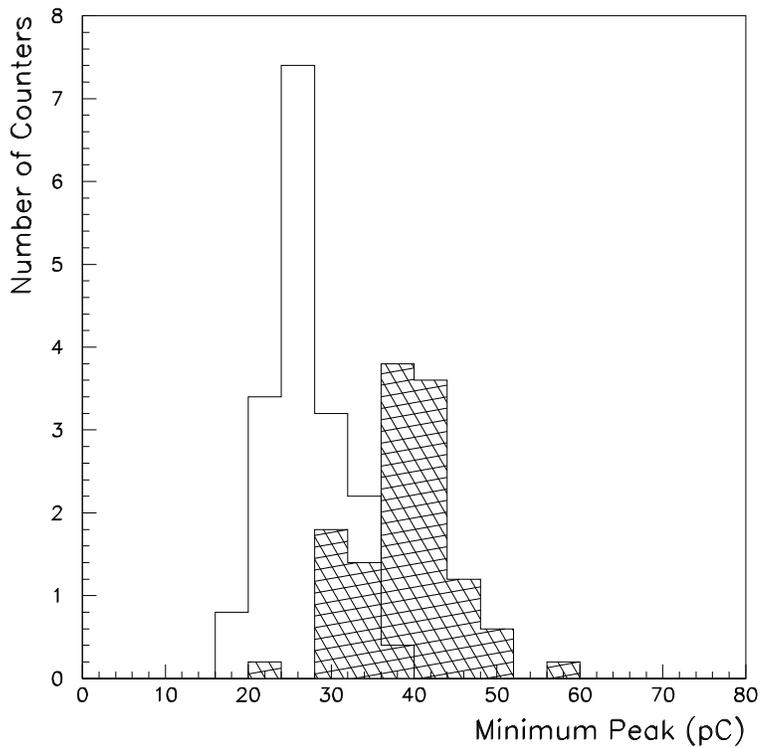}
  \end{tabular}
 \end{center}
\vskip 2.5cm
\caption{The minimum pulse height distribution for the 81''(hatched)
and 113'' counters.}
\label{fig:min_pulse}
\end{figure}

\begin{figure}[ht]
 \begin{center}
    \vskip -8.0cm
  \begin{tabular}{c}
    \epsfxsize = 16cm
        \epsffile{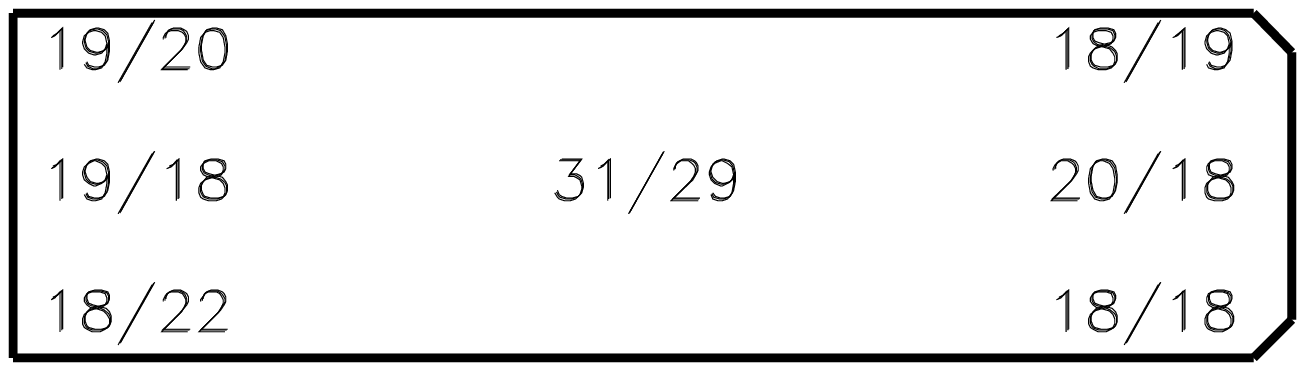}
  \end{tabular}
 \end{center}
\vskip -3.0cm
\caption{The photoelectron yield at different locations along the counter.  The
two numbers are the yield given by each photomultiplier tube on the counter.}
\label{fig:counter_pe}
\end{figure}

Figure~\ref{fig:ave_pulse}
shows a typical integrated pulse height distribution for the 81'' and 113''
counters for cosmic rays passing through a
corner and through the center of the counter.
The distribution of the minimum peak pulse height
from any location on the counter is shown in
Fig.~\ref{fig:min_pulse}.  The counter to counter variation in minimum pulse
height is due mainly to differences in the fiber and
phototube cathode.  
The amount of charged collected even at
the poorest location on the counter is sufficient to maintain a high
efficiency.  
A measurement of the photoelectron yield, on a prototype counter,
gave 29 to 31 photoelectrons at the center of the counter
decreasing to 18 to 22 photoelectrons at the corners of the counter
(see Fig.~\ref{fig:counter_pe}).

    \subsection{Final Testing}
A final test before installation was performed on each completed 
counter using the same testing equipment described in 
section~\ref{sec:install_test}, except that the standard phototubes were
replaced with the final phototubes for that counter.  
The counter was tested at the nominal voltage for each 
tube and then nominal minus 100, 200, and 300~V.
This test
verified that the nominal voltage was $\sim$100\% efficient and gave the
turn on curve for the counter.  This turn on curve allows thresholds
adjustments to be made online without compromising efficiency.
The photomultiplier HV were adjusted so that each tube operated 
approximately 150~V above the knee of the plateau curve. 
\section{Commissioning Tests and Initial Results}
    \subsection{Phased Installation}
The muon scintillation counters cover 6 of the 8 octants in the central muon
detector.  
Octants 0, 1, and 2 were installed at the beginning of the
Run 1b and   octants 3, 4, and 7 were 
installed during the run.  Octant
3 was operational by late February 1994 and octants 4 and 7 were operational by
December 1994 and March 1995 respectively.   Each octant consists of 40
scintillation counters mounted on 5 PDT modules.  
The scintillators mounted on the
top were 113'' long, 
the scintillators on the upper sides were 108'' long, 
and the scintillators on the lower sides were 81.5'' long.
    \subsection{PDT Array Testing}
The intial tests of the counters after installation on the detector involved
powering the arrays and measuring the singles (individual PMT) 
and doubles rates (coincidence between two PMT's) from each
counter.  This verified each counter was in nominal 
working condition and was not damaged
during installation.  Plots of counter hits and hit timing during
data taking verified that the counter system was working.  The time
difference between the two phototubes on the same counter 
was a useful measure of counter performance, with an abnormally large 
width indicating a problem with the counter or TVC.
    \subsection{Tests with 500 nsec and 100 nsec Gate Widths}
In order to time in the counters, data from multiple runs were accumulated
and cuts applied 
to insure that only good reconstructed tracks in the muon system
were kept.  Initially the gate width for the electronics was set at 
500~nsec
to insure that the in time particles would be within the gate.  The initial
plots of the timing for the counters showed that in addition to the in-time
muon peaks there were beam halo and backscatter peaks.
These will be discussed later in 
section~\ref{sec:background}.   After determining the in-time muon peak for
the counters, the gate width was reduced in steps to 200~nsec and finally to
100~nsec.  The 100~nsec width was used as the final gate width for most
of the counters and was not reduced further so that a search for slow
moving particles could be made.   However, the 
gate was reduced to 50~nsec for
a ring of counters on either end of the detector because those counters
had a higher rate of backscatter tracks passing through them.
    \subsection{Results from Scintillator Veto Bit}
\label{sec:veto}
Both phototubes on a scintillation counter were required to fire in order
for a trigger to be generated for that counter in the scintillator trigger
board.   The trigger signals from all 8 scintillation counters on a PDT
were O$\!$Red and the
output was sent to two NIM crates in the collision hall.  The triggers
corresponding to the scintillator counters
of a given octant were
O$\!$Red to produce a
single trigger signal for each of the 6 octants which have
counters.  These signals were stretched and sent to the Moving Counting
House (MCH)
outside of the collision hall.  In the MCH the signals were
fed into logic units in NIM crates.  In addition, coarse centroid trigger (CCT)
pulses from the muon PDT's were also fed into the logic
units.  A coincidence of the scintillator trigger pulse for an octant with the
corresponding CCT pulse defines a SCINT\_CONFIRM trigger term.  An absence of
hits in the scintillator octant and the presence of a CCT pulse for that octant
defines the SCINT\_VETO trigger term.  The final trigger term used for the muon
data was a trigger term SCINT\_VETO\_CONFIRM\_BAR which was formed by the
presence of the SCINT\_VETO and the absence of the SCINT\_CONFIRM.  If
there was a valid scintillator-CCT coincidence in any octant, then a veto
in any other octant was ignored.

\begin{figure}[tb]
\vskip 4.0cm
 \begin{center}
  \begin{tabular}{c}
    \epsfxsize = 10cm \epsffile[100 150 550 550]{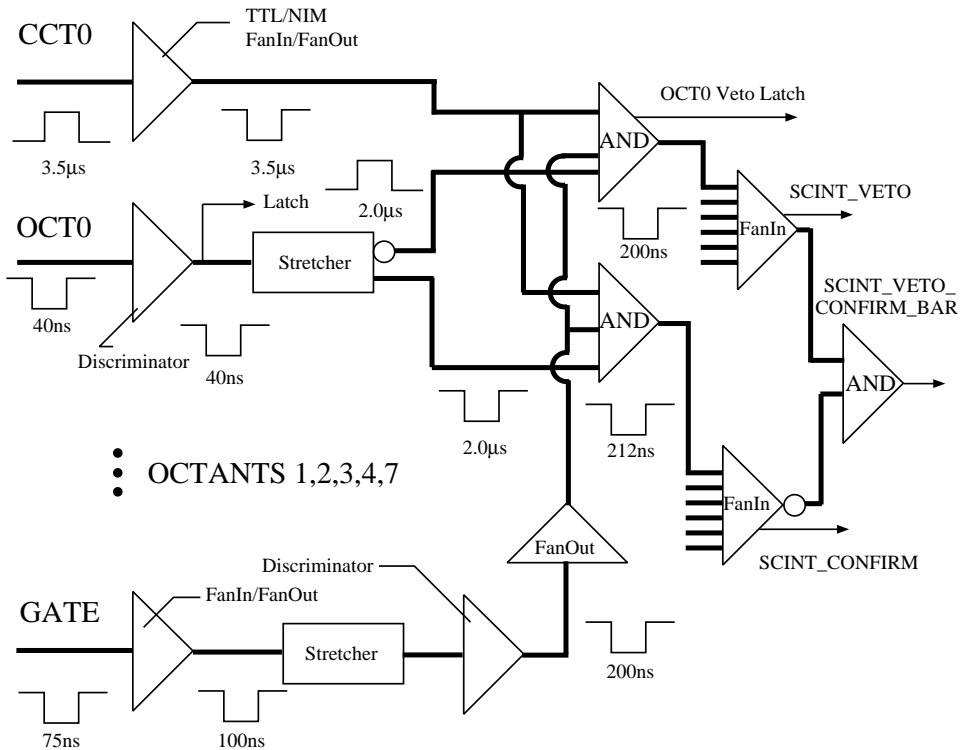}
  \end{tabular}
 \end{center}
\vskip -2.0 cm
\caption{The diagram detailing the logic used in the trigger.}
\label{fig:logic_veto}
\end{figure}

Since the CCT signal was only valid for a short time interval, a gate
was used in
defining the SCINT\_CONFIRM and SCINT\_VETO terms to insure that no spurious CCT
signals would
generate false terms.  The logic diagram for the veto is shown in
Fig.~\ref{fig:logic_veto}.

In addition to
the logical veto terms, 
the scintillator trigger signals and CCT signals for each
octant were recorded in one of the data words for use in monitoring
the Veto.

    \subsection{Monitoring Scintillation Counters}
\begin{figure}[h]
 \begin{center}
  \begin{tabular}{c}
    \epsfxsize = 14cm \epsffile{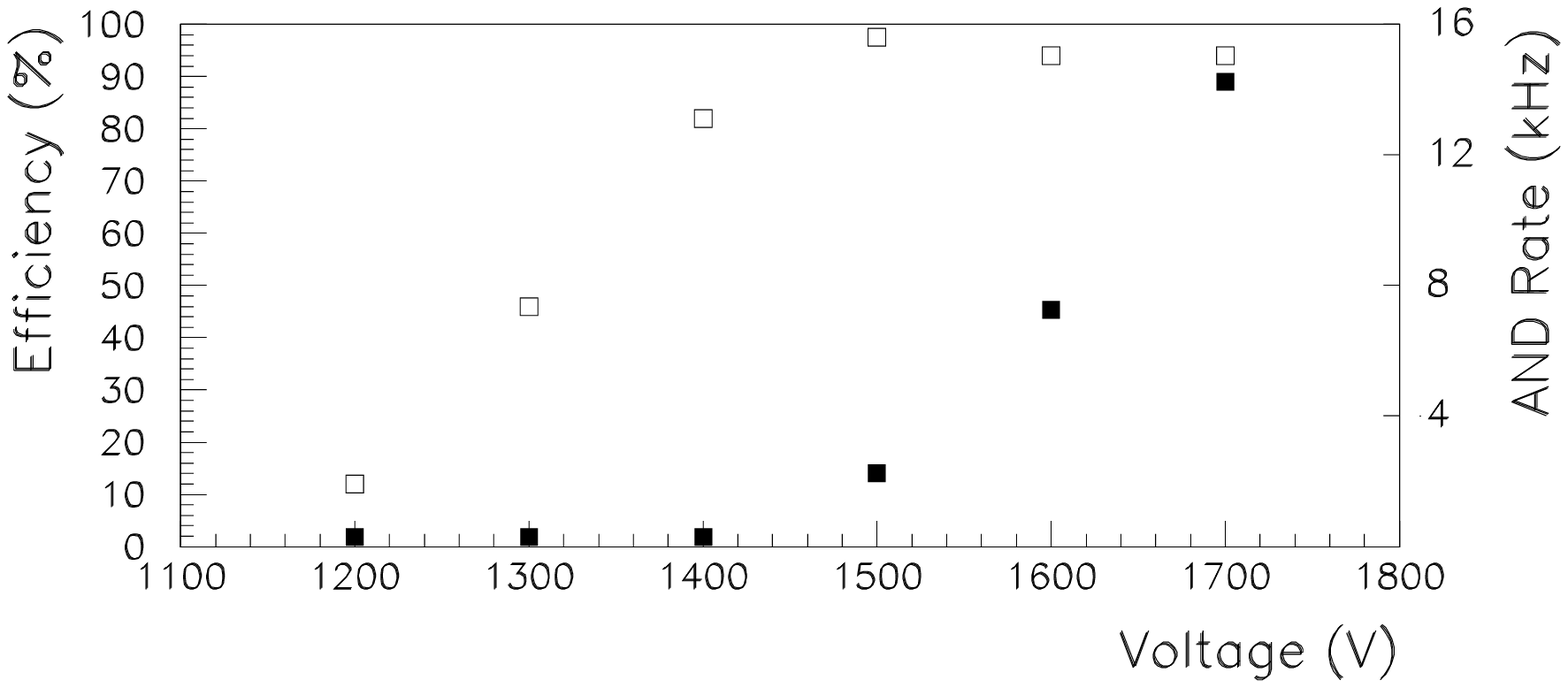}
  \end{tabular}
 \end{center}
\vskip -2.4in
\caption{The efficiency (open squares) and AND rates
(closed squares) for a prototype
counter as a function of the applied High Voltage.}
\label{fig:eff_v_volt}
\end{figure}

The performance of the scintillation counters and the phototube gains were
monitored daily.   During the daily 30 minute ``quiet time'',
where there was
no beam in the accelerator, the AND rates for each of the scintillation counters
were
recorded by a software program which used the trigger board 1553 bus to
selectively disable and enable counters and also place them in the transparent
mode where the discriminator signals are ungated.  The AND rate for fully
efficient counters was dominated by the background radiative
photons from the collision
hall walls and ceiling.  By maintaining the rates at the appropriate level we
could insure that the counters were fully efficient.   Periodically we lowered
the thresholds for counters which were beginning to record a low AND rate to
maintain full efficiency.  The efficiency verses voltage for a prototype
counter placed in the collision hall is shown in
Fig.~\ref{fig:eff_v_volt} and also shown
is the AND rate for the same counter.

\begin{figure}[h]
 \begin{center}
  \begin{tabular}{c}
    \epsfxsize = 8cm \epsffile{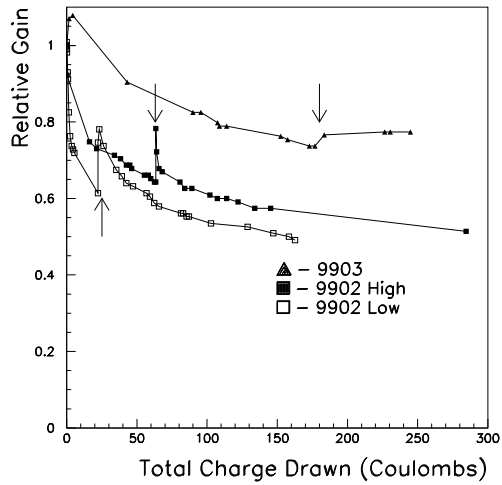}
  \end{tabular}
 \end{center}
\caption{The relative gain as a function of the total charge drawn for a EMI
9903KA and 9902KA phototubes.  The 9902 phototube was run at two voltage values.
The arrows indicate a two week span when the phototube was powered off.}
\label{fig:gain_v_cur}
\end{figure}

As the photomultiplier tubes aged we observed that the gain of some of the
tubes was dropping quite rapidly.
To study this, two sample tubes were tested as a
function of the anode current by using the equipment described in
section~\ref{sec:pmt}.  The
gain verses integrated anode current for a EMI 9902KA tube, which have CsSb
dynodes, is shown in Fig.~\ref{fig:gain_v_cur}.
For this test the EMI 9902KA was run at two voltage values.
Also shown is the response for a EMI 9903KA tube which
has BeCu dynodes.  Other than the rapid initial loss in gain as a function of
integrated anode current, the EMI 9902KA tubes have
operated without problems.  Of the 480
tubes in the system only 3 tubes have failed during the run.

In addition to AND rates the voltages and currents of the HV supplies
which powered the counters were
monitored at the same time.  The ratio of voltage
divided by the current was monitored to check for constancy over time.
This ratio was used to find HV channels which drifted due to moisture being
absorbed into the voltage feedback resistor of the HV supply.  This monitor
was sensitive to drifts  of 10~V or greater.
    \subsection{Counter Performance }
The offline analysis is broken into two parts.  The first is the calibration of
the timing for the counters and determination of the efficiency of each counter.
The second deals with using the scintillator as a tool to improve the purity of
the muon sample.  
        \subsubsection{Timing Calibration}
The time-to-voltage converters of the muon scintillator electronics
were stopped by a
Tevatron clock pulse synchronized to the average beam crossing time.
The voltage was then converted to a digital signal by an
analog to digital converter(ADC).
The time recorded by the electronics includes the time-of-flight of
the muon from the interaction point to the scintillator(T$_{TOF}$),
the propagation time of the light in the fiber(T$_{fiber}$), and
the time delay in the cable and electronics.
A scintillator time ${\mbox T}_{\mu-scint}$, which corresponds to
the time-of-flight of the muon, was calculated from the ADC counts
by using the conversion
\begin{eqnarray}
    {\mbox T}_{\mu-scint} = {\mbox T}_{slope} \cdot
    \left( {\mbox T}_0 - {\mbox T}_{ADC} \right) - {\mbox T}_{fiber}.
\end{eqnarray}
where T$_{ADC}$ is the scintillator time in ADC counts, and 
T$_{slope}$ is a constant, used to convert the ADC counts to time
in nanoseconds.
The T$_0$ was a counter by counter time offset chosen so that the
average T$_{\mu-scint}$ corresponded to the average time of flight
from the interaction point to the scintillation counter.
An initial T$_0$ value was extracted
from the intercept of the calibration fit with the cable delay time.
Collider muons were then used to refine the T$_0$'s by taking the
difference between the expected time and the observed.

\begin{figure}[htb]
 \begin{center}
  \begin{tabular}{cc}
    \epsfxsize = 7cm \epsffile{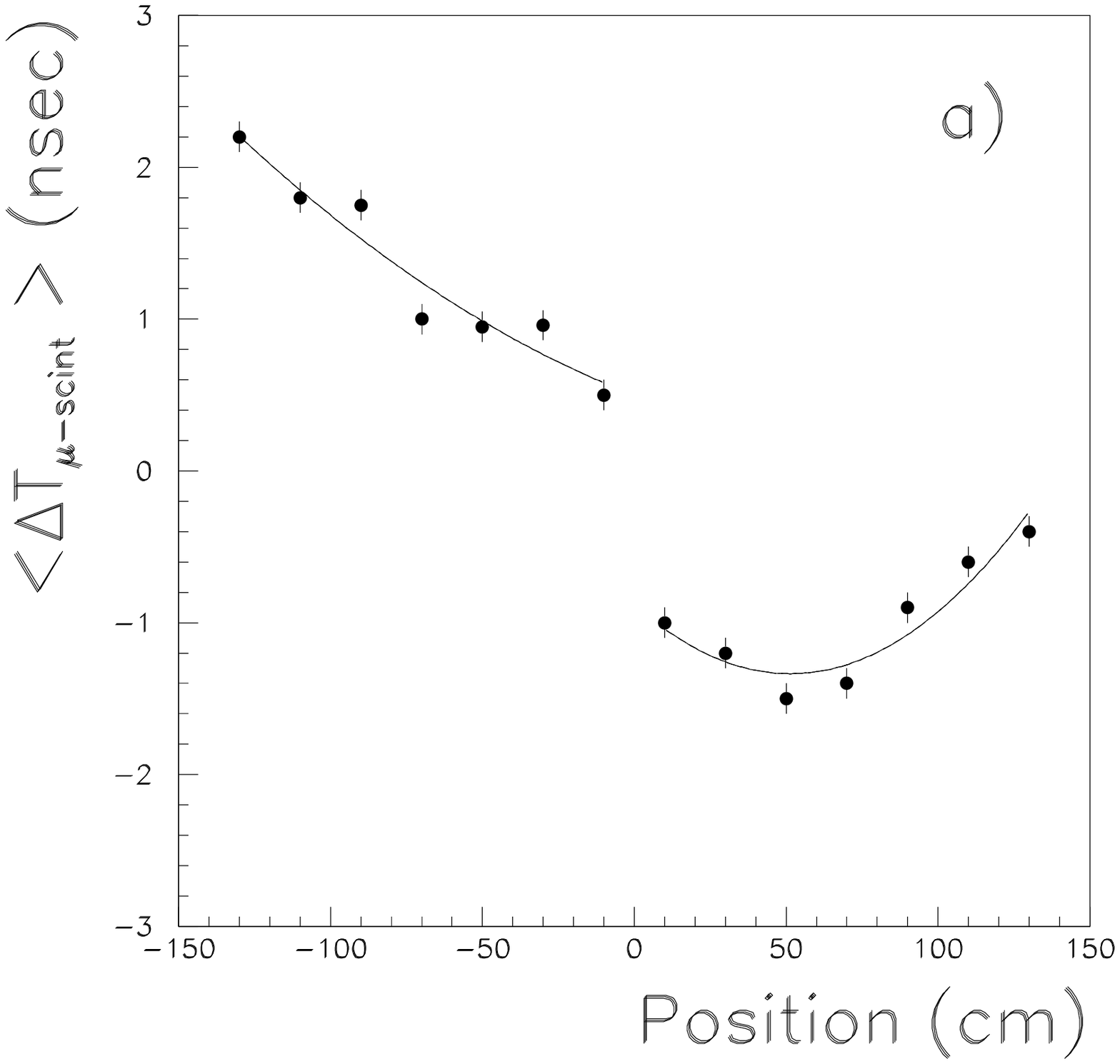} &
    \epsfxsize = 7cm \epsffile{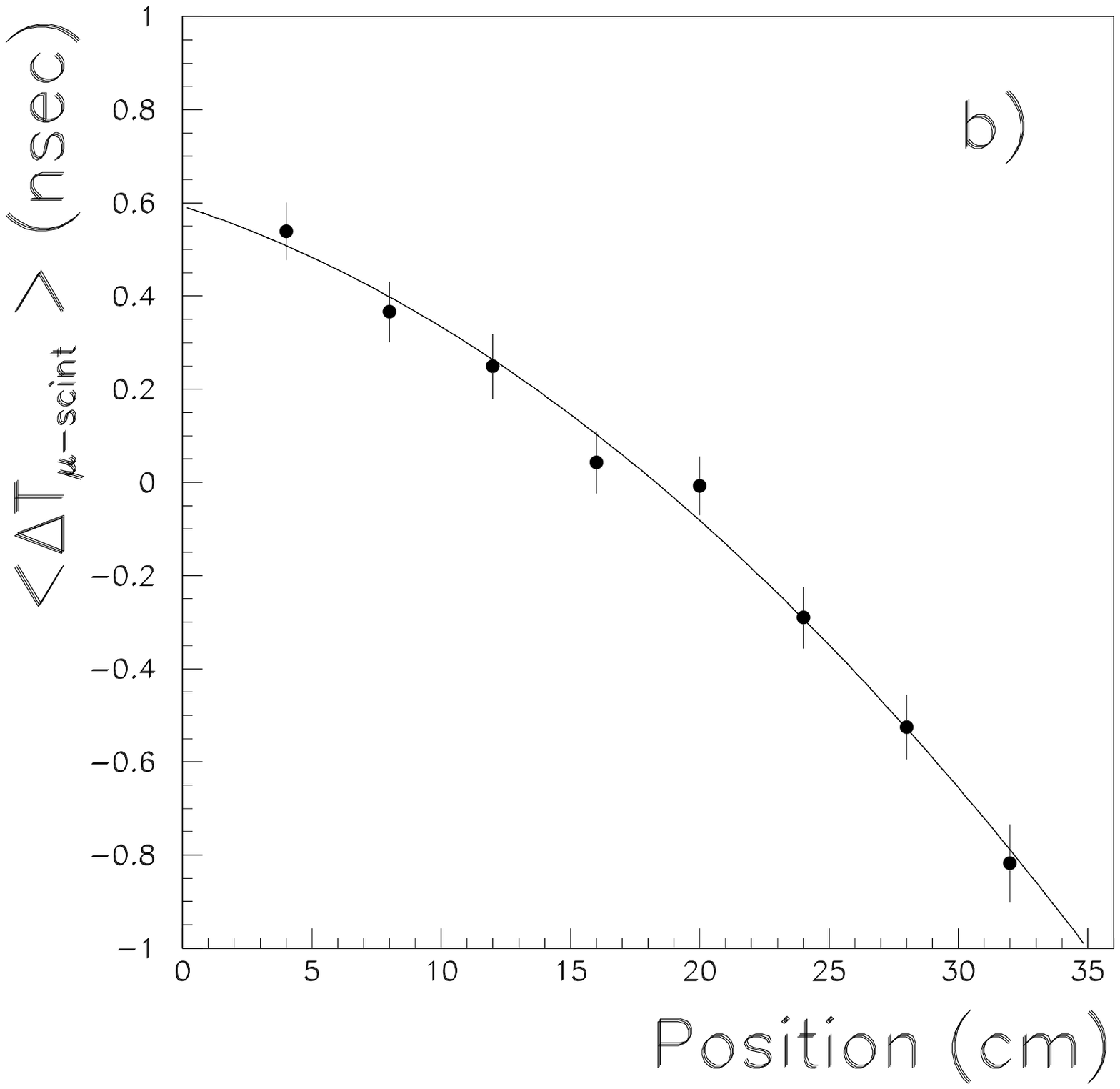} \\
  \end{tabular}
 \end{center}
\caption{The mean $\Delta {\mbox T}_{\mu-scint}$ versus (a) the position of the
muon hit along the length of the counter and (b) along the
width of the counter.  The zero position corresponds to the center
of the counter.}
\label{fig:time_pos}
\end{figure}

Muons detected in the collider data were used to measure 
systematic variations in the timing across the face of the counter.
This was done after the time of flight and a linear
fiber light propagation correction had been applied to
the data.  
These variations result in non-zero values for the time difference
$\Delta {\mbox T}_{\mu-scint}$ between T$_{\mu-scint}$ and the
time-of-flight defined as
\begin{eqnarray}
    \Delta {\mbox T}_{\mu-scint} =
      {\mbox T}_{TOF}
    - {\mbox T}_{\mu-scint}.
\end{eqnarray}
An asymmetry in the timing occurs due to
the different fiber lengths used to readout the two halves of the
scintillator.  The fiber from one end must make a
U-turn to reach the phototube and the fiber from the opposite end does not.
Shown in Fig.~\ref{fig:time_pos} is
$\langle \Delta {\mbox T}_{\mu-scint} \rangle$ versus the position along
the length of the scintillator.  The origin is taken to be the center of the
counter.  The discontinuity is consistent with the fiber from one end of the
counter being about 10'' longer than from the other end.
Figure~\ref{fig:time_pos} also shows
$\langle \Delta {\mbox T}_{\mu-scint} \rangle$ versus the position along the
breadth of the counter where zero is the center of the counter.
The breadth variation is due to longer fiber lengths to phototube for fibers
further from the center of the counter.  
Each of these figures
was fit with a second order polynomial and used as a correction to the time.

\begin{figure}[htb]
 \begin{center}
  \begin{tabular}{c}
    \epsfxsize = 10cm \epsffile[100 150 550 550]{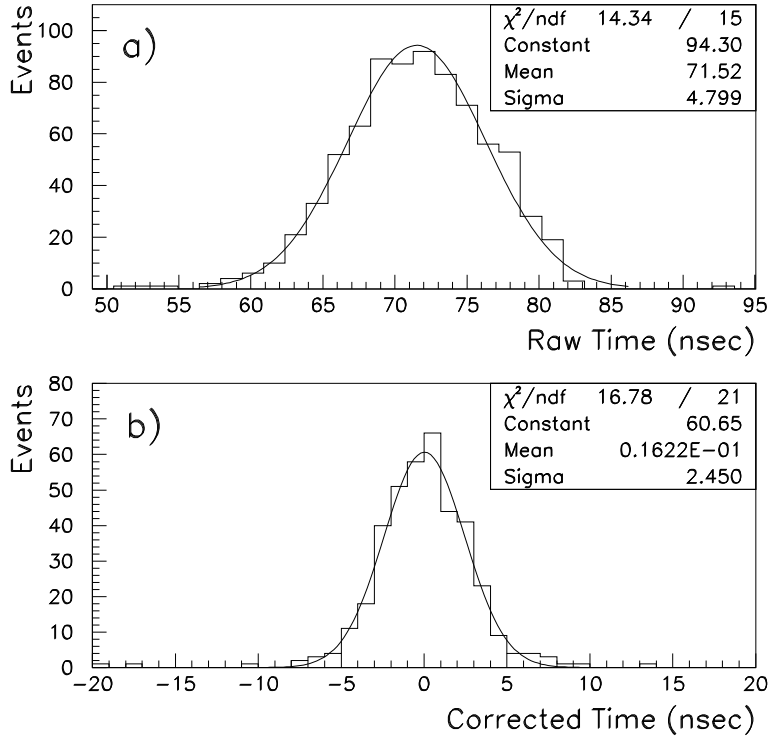}
  \end{tabular}
 \end{center}
\vskip 3.0cm
\caption{(a) The raw time distribution
for an 81'' counter.  And (b) shows the corrected
time. }
\label{fig:time}
\end{figure}

The results of these calibrations are shown in Fig.~\ref{fig:time}.
The top plot shows the raw time distribution from the scintillator and
the bottom
plot is the corrected timing distribution.  The result of the calibration
procedure is an improvement in the resolution by approximately 
a factor of two.  This
allowed the timing cuts on the scintillator to be narrowed 
thereby eliminating additional background events.
        \subsubsection{Counter Efficiency}
The efficiency of the counters has been measured
using
collider muons.
The muons used in the study are required to pass
standard D\O\  quality cuts and point to an octant instrumented
with scintillation counters.  Track quality requirements are necessary to ensure
a good sample of muon tracks.  Poorly reconstructed tracks can fake muon
scintillator inefficiency by reconstructing away from the fired counter.
The efficiency is calculated for each counter as the fraction of reconstructed
tracks pointing to the counter that have a valid hit from
either photomultiplier tube.
The global muon scintillator efficiency from summing over tracks hitting
all counters is $98.4\pm 0.2(stat.)$\%.
We also note that for 98.1\% of the tracks both phototubes fired.

There are a number of reasons why the scintillator counters may be incorrectly
registered as inefficient:
\begin{enumerate}
\item Cosmic ray muons or other out-of-time tracks could fall within the
900~nsec gate of the PDT but not the 100~nsec gate of the scintillator.
\item Cracks between scintillator counters on a chamber.
\item Spurious tracks from reconstruction or other sources.
\item The intercept location of the muon track on the scintillator counters is
reconstructed incorrectly and the fired scintillator is not matched to the
track.
\end{enumerate}
Given these effects we estimate the true counter efficiency as greater than
99\%.  The counter efficiency was also determined to be independent of the track
position in the counter.
        \subsubsection{Background Rejection}
\label{sec:background}
The scintillation counters do an excellent job of rejecting backgrounds to the
muon sample.  The backgrounds are from cosmic rays and accelerator related
effects such as beam halo and backscattered particles.  The beam halo tracks
traverse the detector along with the protons involved in the collision
and arrive early relative to the events associated with the interaction.
The backscatter particles were produced by particles from the interaction, striking
the inner radius of the calorimeter, beam collimators and the accelerator
quadrapole magnets at either end of the detector.   Because of the extra
distance, the backscatter particles arrive later than the events coming directly from 
the interaction.

\begin{figure}[h]
 \begin{center}
  \begin{tabular}{c}
    \epsfxsize = 8cm \epsffile{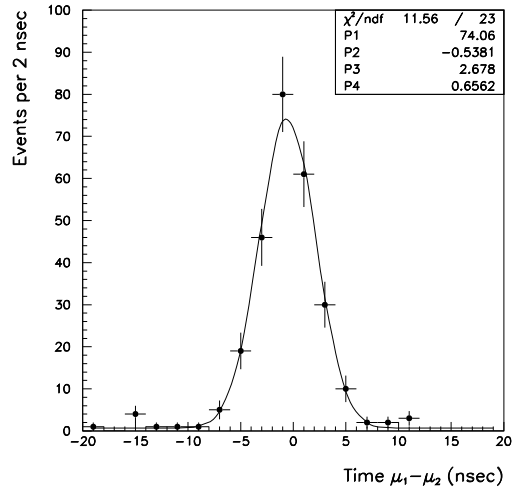}
  \end{tabular}
 \end{center}
\caption{The difference in scintillator times for a sample of dimuon events.}
\label{fig:dimu}
\end{figure}

The fast timing information of the scintillator allow the time-of-flight of
the muon to be used as a discriminator against all three of the mentioned
backgrounds.   Figure~\ref{fig:dimu} shows the difference in the time-of-flight
of flight for a sample dimuon events.  The fit is to a Gaussian plus a
zeroth order polynomial.  The percentage of cosmic muons within a two standard
deviation cut is only 1.4\%.

\begin{figure}[h]
 \begin{center}
  \begin{tabular}{cc}
    \epsfxsize = 7cm \epsffile{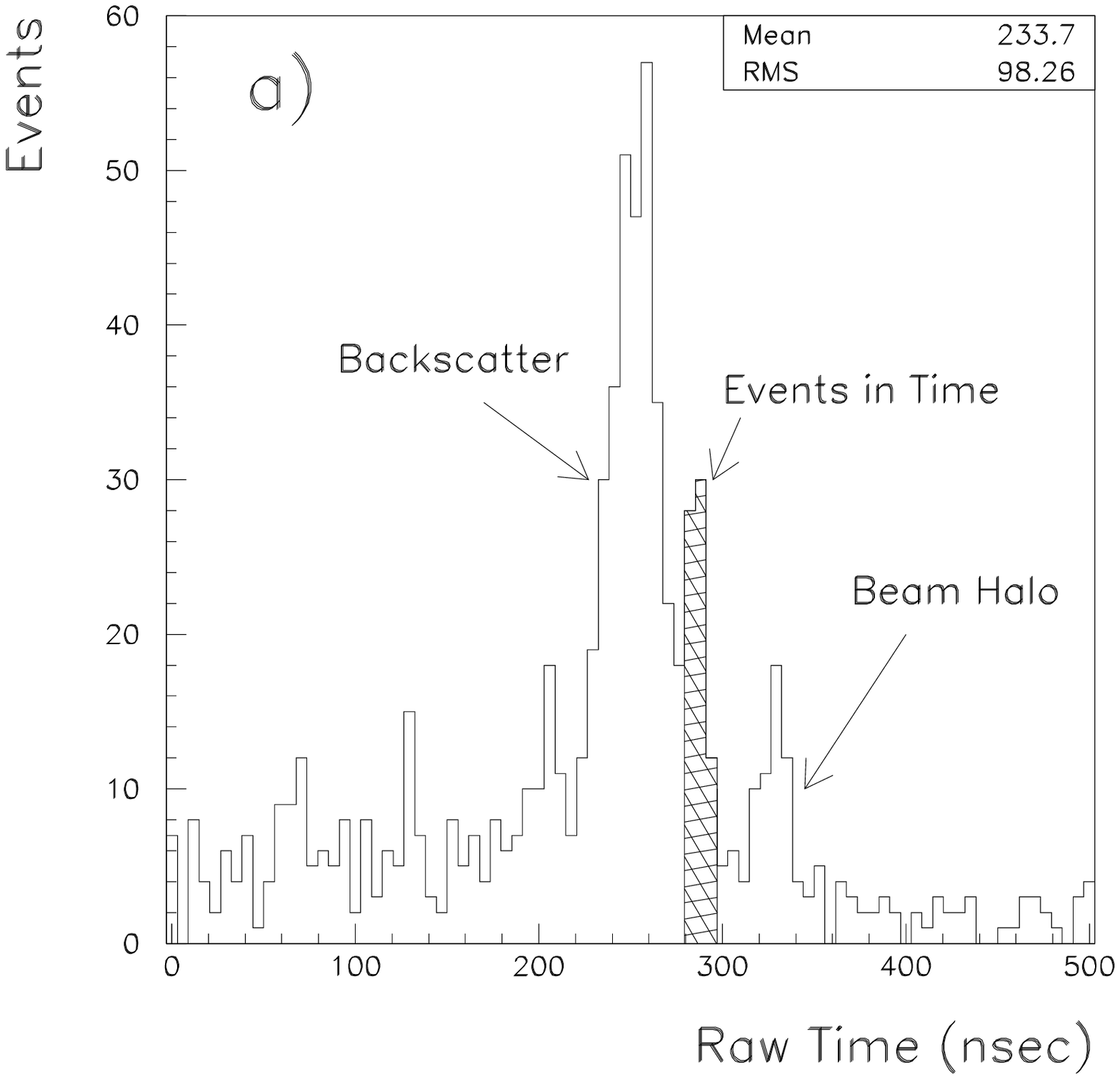} &
    \epsfxsize = 7cm \epsffile{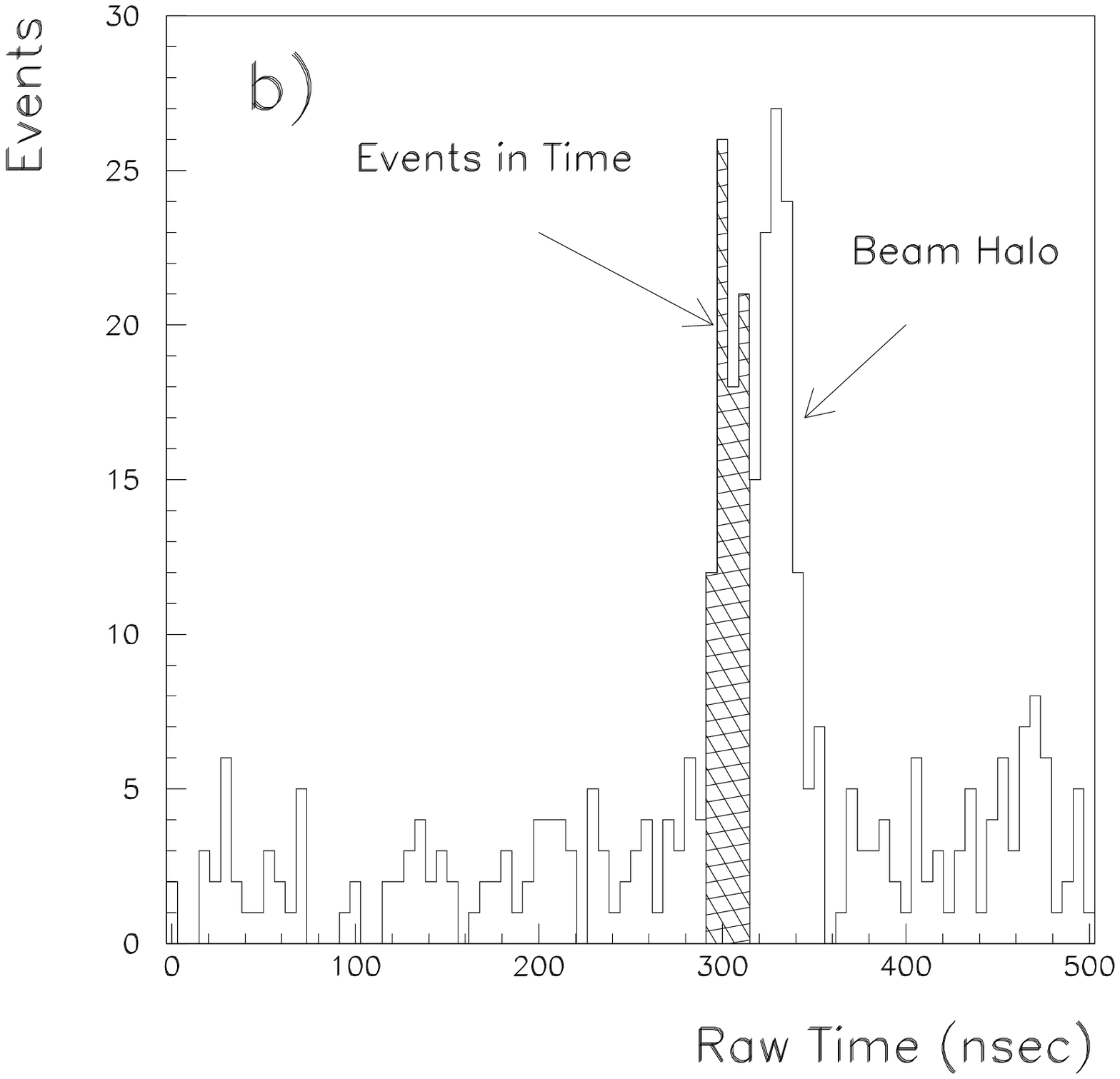} \\
  \end{tabular}
 \end{center}
\caption{(a)The raw time 
distribution of hits where the three peaks are due to beam
halo, backscatter, and in-time events for a counters near one
end of the detector.  (b) And the raw time distribution for a
counters that are near the center of the detector.}
\label{fig:back_scat}
\end{figure}

Figure~\ref{fig:back_scat} shows the raw time distribution
of hits on a scintillator counter near one end of the detector 
and a counter near the center of the detector.   The three
distinct peaks are due to beam halo, backscatter, and in-time events.  The
ring of counters near each end of the detector had the most backscatter
tracks passing through them.  Counters near the center of the detector did
not have such large backscatter peaks as they were shielded from the
backscatter sources by the calorimeter and iron toroid.   The beam halo
peak was reduced substantially after these plots were made by tuning
of the accelerator and by installing new shielding walls just outside the
detector in the accelerator tunnel.  For Run 2, new shielding will be
installed inside the detector which is expected to 
reduce the backscatter by approximately a factor of 50.  

The effect of the scintillator veto on the first level muon trigger was
substantial.  The high p$_{T}$ dimuon trigger was reduced by approximately
a factor of 3 when the veto for the top four octants were added to the trigger.
The low p$_{T}$ dimuon trigger was reduced by a factor of approximately
1.6.
\section{Summary}
The scintillation counters built for the D\O\ Muon System worked very
well during the latest data taking run.  The scintillator and fiber design
chosen performed as expected.  The phototubes chosen had a rather sharp
gain variation during the intial running but were stable after the intial
burn-in period.  The scintillators were effective in rejecting background
from cosmic
rays, out-of-time accelerator muons, and backscatter particles, while
maintaining an efficiency greater than 99\%.
\begin{ack}
We thank the staffs at Fermilab and the collaborating institutions for their
contributions to this work, and acknowledge support from the 
Department of Energy and National Science Foundation (U.S.A.),  
CNPq (Brazil),
Departments of Atomic Energy and Science and Education (India),
and CONACyT (Mexico).  We also thank A. Galyaev who worked with
us on the photocathode measurements.
\end{ack}

\end{document}